# ELUSIVE PREFERRED HOSTS OR NUCLEIC ACID LEVEL SELECTION?

## A commentary on: Evolutionary interpretations of mycobacteriophage biodiversity and host-range through the analysis of codon usage bias (Esposito *et al.* 2016)

Donald R. Forsdyke, Department of Biomedical and Molecular Sciences, Queen's University, Kingston, Ontario, Canada K7L3N6

### Argument for elusive preferred host

While confirming the long held view that "viruses do not closely imitate the use of the [host's] … codon catalogue" (Grantham *et al.,* 1986), it is nevertheless considered a "surprising finding" that "despite having the ability to infect the same host, many mycobacteriophages share little or no genetic similarity" (i.e. similarity in their "GC contents and codon utilization patterns;" Esposito *et al.*, 2016). Arguing correctly that "efficient translation of a phage's proteins within a host is optimized by the phage's ability to match the codon usage patterns of their hosts," the authors conclude that "the preferred host of many mycobacteriophages is not *M. smegmatis*, despite their having been isolated on *M. smegmatis*." Thus, a virus and its elusive preferred hosts would have had similar GC% and codon usages, but the same virus could still infect a less-preferred host (*M. smegmatis*), where the virus-host similarity would be less evident.

### Another evolutionary interpretation

All this rests on the incorrect assumption that efficient translation (protein level selection; Ran *et al.*, 2014) is evolutionarily decisive and cannot be overruled by nucleic acid level selection. Another interpretation is that, early in the diversification into distinctive mycobacteriophage species, prototypic phage lines acquired GC% differences that permitted coinfection of a common host cell by eliminating the recombination-dependent blending of sequences (Forsdyke, 1996). Coinfectants that share a common cytosol either blend or speciate. Thus, selection is primarily at the nucleic acid level and translation efficiency is secondary. So powerful can be the pressure on genomes to avoid recombination that, *in extremis*, a virus that 'needs' to translate more rapidly is 'forced' to encode its own tRNAs tailored to this special need.

### Nucleic acid level selection

Grantham himself had noted that α and β globin mRNAs are translated within the same eukaryotic cell yet have different GC% values and codon usage patterns (Grantham *et al.,* 1986). A simple evolutionary interpretation is that divergence from a prototypic globin gene had been assisted by early-developing GC% differences. These differences had impeded the recombinational blending between the emerging α and β globin genes, which would have reversed the divergence process (Forsdyke, 1996). Likewise, Wyatt (1952) had found that viruses that could coinfect a common host cell diverged *widely* in genome GC% (and hence in codon usage pattern), whereas viruses with different hosts differed *much less* in GC% (and hence in codon usage pattern). Other virus-pair examples include the low GC% HIV and the high GC% HTLV1 that are both hosted by CD4 lymphocytes and are likely derived from the same retroviral ancestor (Forsdyke, 2014; Meyer et al. 2016). The GC% differences may themselves be an expression of more fundamental oligonucleotide differences that bar recombination (Brbić *et al.,* 2015; Forsdyke, 2016). A study that conceded the possibility of nucleic acid level selection (Ran *et al.,* 2014) is cited by Esposito *et al.* (2016), but here the emphasis is on selection on RNA secondary structure rather than at the genome-level (i.e. on *M. smegmatis* DNA).

# References

**Brbić, M., Warnecke, T., Kriško, A. & Supek, F**. **(2015).** Global shifts in genome and proteome composition are very tightly coupled. *Genome Biol Evol* 7, 1519–1532.

**Esposito, L. A., Gupta, S., Streiter, F., Prasad, A. & Dennehy, J. J. (2016).** Evolutionary interpretations of mycobacteriophage biodiversity and host-range through the analysis of codon usage bias. *Microbiol Genom* 2(10), doi: 10.1099/mgen.0.000079.

**Forsdyke, D. R. (1996).** Different biological species "broadcast" their DNAs at different (G+C)% "wavelengths". J Theor Biol 178, 405–417.

**Forsdyke, D. R. (2014).** Implications of HIV RNA structure for recombination, speciation, and the neutralism-selectionism controversy. *Microbes Infect* 16, 96–103.

**Forsdyke, D. R. (2016).** *Evolutionary Bioinformatics*, 3rd edn. New York: Springer.

**Grantham, R., Perrin, P. & Mouchiroud, D. (1986).** Patterns in codon usage of different kinds of species. *Ox Surv Evol Biol* 3, 48–81.

**Meyer, J. R, Dobias, D. T., Medina, S. J., Servilio, L., Gupta, A. & Lenski, R. E. (2016).** Ecological speciation of bacteriophage lambda in allopatry and sympatry. *Science* (in press) doi: 10.1126/science.aai8446

**Ran, W., Kristensen, D. M. & Koonin, E. V. (2014).** Coupling between protein level selection and codon usage optimization in the evolution of bacteria and archaea. *Mbio* 5(2), e00956–14.

**Wyatt, G. R. (1952).** The nucleic acids of some insect viruses. *J Gen Physiol* 36, 201–205.